\documentclass[preprint,aps,draft]{revtex4}
\usepackage{graphicx}
\usepackage{dcolumn}
\usepackage{bm}
\def\br{\begin{eqnarray}}
\def\er{\end{eqnarray}}
\def\be{\begin{equation}}
\def\ee{\end{equation}}

\def\({\left(}
\def\){\right)}

\begin{document}
\title{Ultra peripheral
heavy ion collisions and the energy dependence of the nuclear radius}
\author{C.~G.~Rold\~ao\footnotemark
\footnotetext{e-mail : roldao@ift.unesp.br}}
\author{A. A. Natale\footnotemark
\footnotetext{e-mail : natale@ift.unesp.br}} \affiliation{
Instituto de F\'{\i}sica Te\'orica, Universidade Estadual
Paulista, Rua Pamplona 145, 01405-900, S\~ao Paulo, SP,
Brazil 
}%
\begin{abstract}
To estimate realistic cross sections in ultra peripheral
heavy ion collisions we must remove effects of strong absorption. One method to eliminate
these effects make use of a Glauber model calculation, where the nucleon-nucleon energy
dependent cross sections at small impact parameter are suppressed. In another method we
impose a geometrical cut on the minimal impact parameter of the nuclear collision ($b_{min} > R_1
+ R_2$, where $R_i$ is the radius of ion ``$i$").  In this last case the effect of a possible nuclear radius
dependence  with the energy has not been considered in detail up to now. Here we introduce
this effect showing that for final states with small invariant mass the effect is negligible. However
when the final state has a relatively large invariant mass, e.g. an intermediate mass Higgs
boson, the cross section can decrease up to $50\%$.
\end{abstract}
\maketitle Collisions at relativistic heavy ion colliders like the
Relativistic Heavy Ion Collider RHIC/Brookhaven and the Large
Hadron Collider LHC/CERN (operating in its heavy ion mode) are
mainly devoted to the search of a quark-gluon plasma in central
nuclear reactions. In addition to this important feature of
heavy-ion colliders, ultra peripheral collisions may give rise to
a huge luminosity of photons opening the possibilities of studying
two-photon and other interactions as reviewed in
Refs.\cite{baur1,baur1b,baur2}. In the early papers on peripheral
heavy ion collisions the effect of strong absorption was not taken
into account. The separation of the strong interactions effects
was solved by using impact parameter space methods in Refs.
\cite{gbaur,gbaur2,cahn}. In order to obtain a truly peripheral
photon-photon interaction one has to remove completely the central
collisions, i.e we must enforce that in the cross section
calculation the minimum impact parameter, $b_{min}$, should be
larger than $R_1 + R_2$, where $R_i$ is the nuclear radius of the
ion ``$i$" \cite{gbaur}. The photon distributions can be described
using the equivalent-photon approximation (EPA) with the
requirement of minimum impact parameter (or geometric cut)
discussed above \cite{baur2,cahn}.

The above method is not the only manner to avoid events
where hadronic particle production overshadows the $\gamma -\gamma $
interaction, i.e., events where the nuclei physically
collide. An alternative is to use the Glauber model for
heavy ion collisions \cite{glauber}. It is a semiclassical model
picturing the nuclei moving in a straight path along the collision
direction, and gives the nucleus-nucleus interaction in terms of
the interaction between the constituent nucleons and nuclear
density distributions. If we write the cross section for the collision
of two nucleus $A$ and $B$ as a function of the impact parameter ($b$),
the elastic ($el$) peripheral cross section will be given by
\begin{equation}
\sigma_{el} = \int d^2b [1 - \exp{(-AB \sigma_0 T_{AB}(b)/2)}]^2,
 \label{gla1}
\end{equation}
where $A$ and $B$ are the nucleon numbers, $\sigma_0$ is the total
nucleon-nucleon cross section and
\begin{equation}
T_{AB}(b) = \int   \frac{dQ^2}{(2\pi^2)} F_A(Q^2)
F_B(Q^2)e^{\imath Qb},
 \label{gla2}
\end{equation}
where $F_{A(B)}$ are nuclear form factors. Eq.(1) and the form (2)
for $T_{AB}(b)$ are valid only if one can neglect the finite range
of the nuclear interaction. If at higher energies the total cross
section increases both due to strength and due to the range the
equation for $T_{AB}(b)$ should take this into account. The
exponential factor in Eq.(\ref{gla1}) is the one responsible for
the suppression of the inelastic collisions. The $\sigma_0$ total
nucleon-nucleon cross section that appears in Eq.(\ref{gla1}) is
known to be dependent on the energy. Actually the increase of
hadron-hadron total cross sections have been theoretically
predicted many years ago \cite{inc} and these predictions have
been accurately verified by experiment \cite{pdg}. For instance,
the proton-proton total cross section roughly double as we go from
a few GeV up to the Tevatron energies.

In ultra peripheral heavy ion collisions it is clear how this
energy dependence of the cross section enters in the Glauber approximation.
However the same is not true when we compute the cross sections with
the EPA and the requirement of a minimum impact parameter.
It seems that cross sections in very peripheral  heavy ion collisions calculated within the
Glauber method turn out to be slightly different from the ones computed with the geometric cut \cite{roldao}.

The nuclear radius certainly expands with the increase of the energy in the same way as the proton expands,
and this expansion should be implemented in the geometrical cut calculation of peripheral heavy ion collisions.
As far as we know this effect has not been discussed in detail in the literature, and it is the purpose of this
work to introduce the energy dependence of the nuclear radius in the calculations of peripheral heavy
ion collisions when the geometric cut method is used.

In order to introduce the energy dependence of the nuclear radius in the
calculations of peripheral heavy ion collisions we start discussing
a standard computation of the photon distribution in the ion with the geometric cut method.
The photon distribution in the nucleus can be described using the
Weizs\"{a}cker-Williams approximation (or EPA) in the
impact parameter space. Denoting by $F(x)dx$ the number of photons
carrying a fraction between $x$ and $x+dx$ of the total momentum
of a nucleus of charge $Ze$, we can define the two-photon
luminosity through
\begin{equation}
\frac{dL}{d\tau} = \int ^1 _\tau \frac{dx}{x} F(x) F(\tau/x),
\end{equation}
\noindent
 where $\tau = {\hat s}/s$, $\hat s$ is the square of
the center of mass (c.m.s.) system energy of the two photons and
$s$ of the ion-ion system.
The total cross section of the process
$ ZZ \rightarrow ZZ X $ is
\begin{equation}
\sigma (s) = \int d\tau \frac{dL}{d\tau} \hat \sigma(\hat s),
\label{eq:sigfoton}
\end{equation}
\noindent
 where $ \hat \sigma(\hat s)$ is the cross-section of
the subprocess $\gamma \gamma \rightarrow X$.
There remains only to determine $F(x)$. In the literature there
are several approaches for doing so, and we choose the
conservative and more realistic photon distribution of
Ref.\cite{cahn}. Cahn and Jackson~\cite{cahn}, using a
prescription proposed by Baur~\cite{gbaur}, obtained a photon
distribution which is not factorizable. However, they were able to
give a fit for the differential luminosity which is quite useful
in practical calculations:
\begin{equation}
\frac{dL}{d\tau}=\left(\frac{Z^2 \alpha}{\pi}\right)^2
\frac{16}{3\tau} \xi (z),
 \label{dl1}
\end{equation}
\noindent
 where $z=2MR\sqrt{\tau}$, $M$ is the nucleus mass, $R$
its radius and $\xi(z)$ is given by
\begin{equation}
\xi(z)=\sum_{i=1}^{3} A_{i} e^{-b_{i}z},
 \label{dl2}
\end{equation}
which is a fit resulting from the numerical integration of the
photon distribution, accurate to $2\% $ or better for
$0.05<z<5.0$, and where $A_{1}=1.909$, $A_{2}=12.35$,
$A_{3}=46.28$, $b_{1}=2.566$, $b_{2}=4.948$, and $b_{3}=15.21$.
For $z<0.05$ we use the expression (see Ref.~\cite{cahn})
\begin{equation}
\frac{dL}{d\tau}=\left(\frac{Z^2 \alpha}{\pi}\right)^2
\frac{16}{3\tau}\left(\ln{(\frac{1.234}{z})}\right)^3 .
 \label{e5}
\end{equation}

Eq.(3) is written in a factorised form, which of course is valid
only if one neglects the exclusion of central collisions into
account. Therefore Eq.(3) is not the most general form,
\cite{cahn,baur2}, and the same is true for Eq.(5). The
calculation assumes that the same radius R is used for both ions
$b_{min}=2R$ but also to have a cutoff for the individual impact
parameter $b_1$ and $b_2$ (which either is necessary to eliminate
final state interaction, or which takes into account the form
factor effects, that is, the decrease of the charge inside the
nucleus). Especially when looking, for instance, an intermediate
mass Higgs boson production or other non-strongly interacting
particles there is no reason to assume that the size of the
individual cutoff radii for $b_1$ and $b_2$ scales in the same way
as $b_{min}$. Therefore the calculation overestimates the
dependence on R a bit.

The condition for realistic peripheral collisions ($b_{min} > R_1
+ R_2$) is present in the photon distributions showed above.
To obtain the above equations an elastic Gaussian form factor and an energy independent
nuclear radius giving by $R_{ion}=1.2A^{1/3}$ fm  have been used. A more accurate Woods-Saxon distribution
for symmetric nuclei would produce some small deviations, but for our purposes the expressions for the
luminosity described above are enough. However the expression for the nuclear radius
is exactly the one we believe that should be changed by its energy dependent expression, and
the problem is to have a phenomenologically sensible expression for the nuclear
radius increase with the energy.

In the heavy ions colliders nucleus like Au and Pb will collide
with a great amount of energy, going from 200 GeV/nucleon (Au at RHIC) up to 5.5 TeV/nucleon
(Pb at LHC), and the ultra peripheral collisions can be computed
with the help of the photon distribution described above.
If the ion radius increase with the energy, the value corresponding to
$b_{min}$ will also become greater, and consequently the cross
section must decrease. This is easily seen in the many examples calculated in the literature
where the cross section for a given process is concentrated at some moderate impact parameter
and decreases when b increases. Of course the Lorentz factor is also important to determine
this behavior. Therefore, if we introduce the energy dependence in the nuclear radius we expect lower
rates for a given process than those obtained in the usual calculations, and this effect, even if it is small,
could be important if we have a high precision measurement.

The authors of Ref. \cite{Barroso:vm} modelled the particle
production process in ultrarelativistic heavy ion collisions in
terms of an effective scalar field produced by the colliding
objects, in their work they showed that the nuclear cross sections increase with the
energy due to a logarithmic increase of the nuclear radius with the energy.
We shall use this reference to obtain a relation between the nuclear radius and the
incident energy that is the following:
\begin{equation}
R^2_H(s) = 1 + 2 \frac{d}{R^\prime} \gamma_E + (\delta +1)
\frac{d}{R^\prime} \ln \left( \frac{A\sqrt{s}}{\varepsilon_0}
\right).
 \label{rahs}
\end{equation}
$R^\prime = R_P + R_T \simeq 2.4 A^{1/3}$ fm ($R_P (R_T)$ means
projectile (target) and we assume $R_P = R_T=R_{ion}$), $\sqrt{s}$
is the energy of the projectile nucleus in the laboratory rest
frame. The nuclear density for a nucleus $A$ at distance $x$ from
its center is modelled by a Woods-Saxon distribution for symmetric
nuclei,
\begin{equation}
\rho _{WS}(x) = \frac{\rho_0}{\left(1+ \exp \left[
\frac{(x-R_{ion})}{d} \right]\right)}.
 \label{eq:woods}
\end{equation}
where $d = 0.549$ fm, and $\rho_0$ can be found when the
Wood-Saxon density is normalized by the condition $\int d^3x
\rho(x) = A $. And $\varepsilon_0$ is equal to
\begin{equation}
\varepsilon_0 = M_Z d\left[ \frac{16}{\pi^2g^2\rho_0^4}
\frac{R^\prime}{d^3} \frac{(R_T R_P)^{(\delta-7)/4}}{\Gamma^2
\left( \frac{\delta+1}{4} \right)}
 \right]^{\frac{2}{\delta+1}}.
 \nonumber
\end{equation}
$M_Z$ is the nuclear mass. The
coupling constant $g$ and the parameter for the mass spectrum
$\delta$ were estimated in Ref.\cite{Barroso:vm} and they are
equal to $\delta = -0.56$ and $g = 3.62$ fm$^{(7 + \delta)/2}$.

The radius  $R^2_H(s)$ appearing in Eq.(\ref{rahs}) at small energies gives
a nuclear radius larger than  $R_{ion}$, for this reason we have
assumed the following normalization
\begin{equation}
R^2(s) = \frac{R_H^2(s) R^2_{ion}}{R_H^2(s=M_Z^2)},
 \label{eq:Rs}
 \end{equation}
where $R_{ion} = 1.2 A ^{1/3}$ fm. With this normalization
 factor we assure that when the ion energy is equal to its
 mass, the nuclear radius will be equal to $R_{ion}$. It is the radius given
 by Eq.(\ref{eq:Rs}) that should be used in Eqs.(\ref{dl1} -
 \ref{e5}). Typical values for $R_{ion}$ and $R(s)$ are showed in
 Table \ref{tab:Raio}.

\begin{table}
\begin{tabular}{c c c c}
  \hline \hline
$\,\,\,$Ion $\,\,\,$&$\,\,\,$ $\sqrt{s}$ $\,\,\,$ &$\,\,\,$ $R_{ion}$$\,\,\,$ &$\,\,\,$ $R(s)$$\,\,\,$\\
  \hline
Au  & 0.2 & 6.98 & 7.29 \\
Ca & 7.2  & 4.10 & 4.75 \\
Pb & 5.5  & 7.11 & 7.61 \\
  \hline \hline
\end{tabular}
\caption{ Values for $R_{ion}$ and $R(s)$, Eq.(11), in fm. The
energies ($\sqrt{s}$) are in TeV/nucleon. }
 \label{tab:Raio}
\end{table}

 To show the effects of the nuclear radius dependence on the energy we
computed production of leptons pairs (muons and
taus) and resonances formed by the  photon-photon fusion. In the resonance
case we considered the $\eta_c$ meson and an intermediate mass Higgs boson with a mass equal to 115 GeV.
We computed the cross sections for two cases: In one the nuclear radius is energy
independent (and equal to $R_{ion}$). In the second case the radius obeys
Eq.(\ref{eq:Rs}).

For an invariant mass of the photon pair above the threshold
$\sqrt{\hat{s}} > 2 m_l$, a lepton pair can be produced in
two-photon collisions ($\gamma \gamma \rightarrow l^+ l^-$) and
the lowest order QED cross section for this subprocess is given by
\cite{baur2}
\begin{equation}
\sigma (\gamma \gamma \rightarrow l^+ l^-) = \frac{4 \pi
\alpha^2}{\hat{s}} \beta_l \left[ \frac{(3 - \beta^4_l)}{2\beta_l}
\ln \left( \frac{1+\beta_l}{1-\beta_l} \right) - 2 + \beta^2_l
\right], \nonumber
\end{equation}
where $\beta_l = \sqrt{1 - 4 m_l^2/\hat{s}}$ is the velocity of
the pair in the $\gamma \gamma$ rest frame, $m_l$ is the lepton
mass, $\sqrt{\hat{s}}$ is the c.m. system energy of the two
photons and $\alpha $ is the fine-structure constant. Using this
elementary cross section in Eq.(\ref{eq:sigfoton}) we obtained the
rates shown in Table \ref{tab:Rmuon}. The calculation was
performed for three different ions with different beam energies,
the one of RHIC (Au) and the ones expected at LHC (Ca and Pb). The
cross sections were integrated in a bin of energy equal to $1 <
\sqrt{\hat{s} }< 10$ GeV.  The third column of Table
\ref{tab:Rmuon} shows the cross section computed with a constant
nuclear radius and the fourth column the one with the energy
dependent radius.  For the three different ions the cross sections
decrease when we  consider the energy dependent radius described
by the Eq.(\ref{eq:Rs}). In all the cases the decrease is smaller
than $10\%$ and is negligible considering the theoretical and
experimental uncertainties involved in the problem.

\begin{table}
\begin{tabular}{c c c c c}
  \hline \hline
$\,\,\,$Ion $\,\,\,$&$\,\,\,$ $\sqrt{s}$ $\,\,\,$ &$\,\,\,$ $\sigma_{R_{ion}}$$\,\,\,$ &$\,\,\,$ $\sigma_{R(s)}$    $\,\,\,$         & Ratio $\,\,\,$\\
  \hline
Au  & 0.2 & 2.127 & 1.947 & 1.09 \\
Ca & 7.2  & 0.643 & 0.588 & 1.09  \\
Pb & 5.5  & 106.4 & 101.3  & 1.05 \\
  \hline \hline
\end{tabular}
\caption{ Cross sections of the process $ZZ \gamma \gamma \rightarrow ZZ \mu^+ \mu^-$.
The cross sections $\sigma_{R_i }$ ($\sigma_{R(s)}$)  given in the third (fourth)
column are the ones computed with the energy independent
(dependent) radius. The last column shows the ratio between
the third and fourth columns. The cross sections are in mbarn and
the energies ($\sqrt{s}$) are in TeV/nucleon. }
 \label{tab:Rmuon}
\end{table}
In Table \ref{tab:Rtau} it is possible to see the results when the
subprocess analyzed is $\gamma \gamma \rightarrow \tau^+ \tau^-$
with $2 m_{\tau} <\sqrt{ \hat{s}} < 10 \hbox{GeV}$. The general
behavior of the $\tau$ pair production cross sections is very
similar to the one observed previously in  Table \ref{tab:Rmuon}.
Of course, the cross sections for producing tau pairs are smaller.
However the collision of Au-Au and Ca-Ca are now more sensitive to
the energy dependence of the nuclear radius, producing an effect
larger than $10\%$. The rates for tau pairs production in Pb
collision with a c.m. energy equal to 5.5 TeV/nucleon,  with and
without energy dependence in the ion radius are not so different.
As we shall discuss later the larger cut that we perform in the
impact parameter when we consider the energy dependent radius
removes photons of larger energy. Therefore for final states with
larger invariant masses we may expect a larger effect.

\begin{table}
\begin{tabular}{c c c c c}
  \hline \hline
$\,\,\,$Ion $\,\,\,$&$\,\,\,$ $\sqrt{s}$ $\,\,\,$ &$\,\,\,$
$\sigma_{R_{ion}}$$\,\,\,$ &$\,\,\,$ $\sigma_{R(s)}$    $\,\,\,$         & Ratio $\,\,\,$\\
  \hline
Au  & 0.2 & $6.972 \times 10^{-4}$ & $5.727 \times 10^{-4}$ & 1.22 \\
Ca & 7.2  & $5.176 \times 10^{-3}$ & $4.604 \times 10^{-3}$ & 1.12   \\
Pb & 5.5  & 0.759 & 0.718 & 1.05 \\
  \hline \hline
\end{tabular}
\caption{ The same as in Table \ref{tab:Rmuon}, but for
the subprocess $\gamma \gamma \rightarrow \tau^+ \tau^-$. }
 \label{tab:Rtau}
\end{table}

Let us now consider the case of heavy resonances. To estimate
the production of one resonance $R$ formed by a
photon-photon fusion in peripheral heavy ion collisions we use the following
elementary cross section in Eq.(\ref{eq:sigfoton}),
\begin{equation}
\sigma (\gamma \gamma \rightarrow R) = \frac{8 \pi^2}{M_{R} s}
\Gamma (R \rightarrow \gamma \gamma) \delta \left( \tau -
\frac{M^2_{R}}{s} \right)
\end{equation}
where $M_R$ is the resonance mass and $\Gamma (R \rightarrow
\gamma \gamma)$ its decay width into two photons. In Table
\ref{tab:Reta} we show the results obtained for two-photon
production of $\eta_c$ in peripheral heavy ion collisions with
$M_{\eta_c} = 2.979$ GeV and $\Gamma (\eta_c \rightarrow \gamma
\gamma) = 6.6$ keV. The ratio of the cross sections considering
the two scenarios are 1.16 and 1.11 for Au and Ca ions,
respectively, and 1.06 for the Pb ion.  Finally, in Table
\ref{tab:RHiggs} it can be observed the values corresponding to
the subprocess $\gamma \gamma \rightarrow H$ with $M_H = 115$ GeV,
where we used the Higgs boson two-photon decay width found in Ref.
\cite{Higgs}. We do not show the result for RHIC energies because
it is too small. The values of Table \ref{tab:RHiggs} indicate
that the production cross sections for both ions are strongly
affected by the inclusion of a radius described by
Eq.(\ref{eq:Rs}). In the case of  Ca collision with a c.m. energy
of 7.2 TeV/nucleon the cross sections decrease nearly to half of
the value obtained in the case of a energy independent radius. The
situation is less drastic for the Pb ion with $\sqrt{s}$ = 5.5
TeV/nucleon, but the ratio is still large ($= \,\, 1.34$). This is
the only situation that we investigated where the Pb collision is
clearly sensitive to the use of Eq.(\ref{eq:Rs})  (or to the
energy dependence of the nuclear radius).

\begin{table}
\begin{tabular}{c c c c c}
  \hline \hline
$\,\,\,$Ion $\,\,\,$&$\,\,\,$ $\sqrt{s}$ $\,\,\,$ &$\,\,\,$ $\sigma_{R_{ion}}$$\,\,\,$ &$\,\,\,$ $\sigma_{R(s)}$    $\,\,\,$         & Ratio $\,\,\,$\\
  \hline
Au  & 0.2 & $2.147 \times 10^{-3}$ & $1.846 \times 10^{-3}$ & 1.16 \\
Ca & 7.2  & $2.897 \times 10^{-3}$ & $2.614 \times 10^{-3}$ & 1.11   \\
Pb & 5.5  & 0.437 & 0.413 & 1.06 \\
  \hline \hline
\end{tabular}
\caption{ The same as in Table \ref{tab:Rmuon}, but for
the subprocess $\gamma \gamma \rightarrow \eta_c$. }
 \label{tab:Reta}
\end{table}

The fact  that a sharp cutoff in impact parameter space at  $b_{min}$ should be replaced by a
smooth one was already discussed in Ref.\cite{baur1b}. Comparing the Glauber model calculation
with the one with a sharp cutoff we could expect significant deviations present at the upper end of
the invariant mass distribution. Looking at the photon luminosity we see that only the smallest
impact parameter contribute significantly to the events with large invariant masses.  Imposing the cut on
$b_{min}$ but now with the radius described by Eq.(\ref{eq:Rs}) we obtain a more realistic
calculation of the very peripheral heavy ion collisions.

\begin{table}
\begin{tabular}{c c c c c}
  \hline \hline
 $\,\,\,$Ion $\,\,\,$&$\,\,\,$ $\sqrt{s}$ $\,\,\,$ &$\,\,\,$ $\sigma_{R_{ion}}$$\,\,\,$ &$\,\,\,$ $\sigma_{R(s)}$    $\,\,\,$         & Ratio $\,\,\,$\\
  \hline
Ca & 7.2  & $9.970 \times 10^{-10}$ & $6.789 \times 10^{-10}$ & 1.47   \\
Pb & 5.5  & $1.854 \times 10^{-8}$ & $1.387 \times 10^{-8}$ & 1.34 \\
  \hline \hline
\end{tabular}
\caption{ The same as in Table \ref{tab:Rmuon}, but for
the subprocess $\gamma \gamma \rightarrow H $. }
 \label{tab:RHiggs}
\end{table}

In conclusion, we discussed the two different ways to compute cross sections for ultra peripheral
heavy ion collisions. In the Glauber method it is quite clear how the increase with the energy
of the nucleon-nucleon cross section enters in the calculation. In the calculation with the geometrical
cut imposed on the impact parameter, the nucleon, as well as the nuclear, radius expansion with
the energy was not introduced up to now. It was noticed in Ref.\cite{roldao} that there was a difference
between the two methods. The difference was small and had some dependence on the invariant mass
of the final states. The work of Ref. \cite{Barroso:vm} prescribe a very precise way to introduce
the nuclear radius dependence with the energy.

We believe that the estimative of the cross sections in
ultra peripheral collisions with the geometrical cut method just changing the radius independent of
the energy by the one dependent of the energy will give realistic predictions for any invariant mass
of the final state. The effect is of order of $50\%$ for an intermediate mass Higgs boson.
Turning the problem the other way around we may also say that if the ultra peripheral collisions are
measured with high precision, we may have a new way to study the increase of the nuclear radius with the energy.
To do so we just have to measure the cross sections for very known final states with small and large invariant
masses with high precision, there should be a decrease of the cross sections  as a function of the invariant
mass as we go to larger and larger energies.

{\bf Note added:} Some comments on the effects discussed in this
work were also made by S. Klein and  J. Nystrand in
\cite{Nystrand:1998hw}, where the Fig.3 gives the reduction in
gamma-gamma luminosity (for gold at RHIC) for a Glauber
calculation of hadronic interactions compared to the one with
geometrical cut.

\section*{Acknowledgments}
We are grateful to Y. Hama  for discussions.
This research was supported by the Conselho Nacional de Desenvolvimento Cient\'{\i}fico e Tecnol\'ogico (CNPq)
(AAN) and by Funda\c c\~ao de Amparo \`a Pesquisa do Estado de S\~ao Paulo (FAPESP) (CGR).


\begin {thebibliography}{99}
\bibitem{baur1} C. A. Bertulani and G. Baur, Phys. Rep.
{\bf 163}, 299 (1988).

\bibitem{baur1b} G. Baur, J. Phys. {\bf G24}, 1657 (1998).

\bibitem{baur2} G. Baur, K. Hencken, D. Trautmann,
S. Sadovsky and Y. Kharlov, Phys. Rep. {\bf 364}, 359 (2002).

\bibitem{gbaur} G. Baur, in CBPF Intern. Workshop on
Relativistic Aspects of Nuclear Physics, (Rio de Janeiro, 1989),
edited by T. Kodama et al. (World Scientific, Singapore, 1990), p. 127.

\bibitem{gbaur2} G. Baur and L. G. Ferreira Filho, Nucl. Phys. {\bf A518},
786 (1990).

\bibitem{cahn} R.~N.~Cahn and J.~D.~Jackson,
Phys.\ Rev.\ {\bf D42}, 3690 (1990).

\bibitem{glauber} R. J. Glauber, {\it Lectures on Theoretical Physics}
(Inter-Science, New York, 1959) Vol.I.

\bibitem{inc} H. Cheng and T. T. Wu, Phys. Rev. Lett. {\bf 24}, 1456 (1970);
C. Bourrely, J. Soffer and T. T. Wu, Phys. Rev. Lett. {\bf 54}, 757 (1985);
H. Cheng and T. T. Wu, Expanding protons: Scattering at high energies (MIT
Press, Cambridge, MA, 1987).

\bibitem{pdg} K. Hagiwara et al., Phys.\ Rev.\ D {\bf 66} 010001 (2002)

\bibitem{roldao} C. G. Rold\~ao, Ph.D. Thesis (unpublished); S. Klein,
private communication.

\bibitem{Barroso:vm} M.~F.~Barroso, T.~Kodama and Y.~Hama,
Phys.\ Rev.\ C {\bf 53}, 501 (1996);
 T.~Kodama, S.~J.~Duarte, C.~E.~Aguiar, A.~N.~Aleixo,
M.~F.~Barroso, R.~Donangelo and J.~L.~Neto,
Nucl.\ Phys.\ A {\bf 523}, 640 (1991).

\bibitem{Higgs} R.~Bates and J.~N.~Ng,
Phys.\ Rev.\ D {\bf 33}, 657 (1986).

\bibitem{Nystrand:1998hw}
J.~Nystrand and S.~Klein  [The STAR Collaboration],
arXiv:nucl-ex/9811007.

\end {thebibliography}
\end{document}